
\input harvmac

\def\Titleh#1#2{\nopagenumbers\abstractfont\hsize=\hstitle\rightline{#1}%
\vskip .5in\centerline{\titlefont #2}\abstractfont\vskip .5in\pageno=0}

\def\nextline{\unskip\nobreak\hfill\break}

\catcode`\@=11 

\def\lsim{\mathrel{\mathpalette\@versim<}}
\def\gsim{\mathrel{\mathpalette\@versim>}}
\def\@versim#1#2{\vcenter{\offinterlineskip
    \ialign{$\m@th#1\hfil##\hfil$\crcr#2\crcr\sim\crcr } }}
\def\boxit#1{\vbox{\hrule\hbox{\vrule\kern3pt
      \vbox{\kern3pt#1\kern3pt}\kern3pt\vrule}\hrule}}

\def\t1{{\tilde 1}}

\def\DVN{D. V. Nanopoulos}

\def\GeV{\,{\rm GeV}}
\def\TeV{\,{\rm TeV}}

\def\st{\sin^2\theta_W(m_{Z})}
\def\ae{\alpha_{em}(m_{Z})}
\def\as{\alpha_3(m_Z)}
\def\mnsg{minimal non-supersymmetric $SU(5)$}
\def\msg{minimal supersymmetric $SU(5)$}
\def\msfg{minimal supersymmetric Flipped $SU(5)\times U(1)$}

\def\NPB#1#2#3{Nucl. Phys. B {\bf#1} (19#2) #3}
\def\PLB#1#2#3{Phys. Lett. B {\bf#1} (19#2) #3}

\def\PRD#1#2#3{Phys. Rev. D {\bf#1} (19#2) #3}
\def\PRL#1#2#3{Phys. Rev. Lett. {\bf#1} (19#2) #3}

\def\MODA#1#2#3{Mod. Phys. Lett. A {\bf#1} (19#2) #3}

\def\TAMU#1{Texas A \& M University preprint CTP-TAMU-#1}

\nref\GeorgiI{H. Georgi and S.L. Glashow, \PRL{32}{74}{438}.}
\nref\SU{S. Dimopoulos and H. Georgi, \NPB{193}{81}{150}; \nextline
N. Sakai, Z. Phys. {\bf{C11}} (1982) 153;\nextline
A. Chamseddine, R. Arnowitt and P. Nath, \PRL{105}{82}{970}.}
\nref\EKNI{J. Ellis, S. Kelley, and \DVN, \PLB{249}{90}{441}.}
\nref\AMALDI{U. Amaldi, W. De Boer, and H. Furstenau, \PLB{260}{91}{447}.}
\nref\EKNII{J. Ellis, S. Kelley, and \DVN, \PLB{260}{91}{131}.}
\nref\PDECAY{B. Campbell, J. Ellis and D.V. Nanopoulos,
\PLB{141}{84}{229}; \nextline
R. Arnowitt and P. Nath, \PRD{38}{88}{1479}.}
\nref\LPN{H. Pois, J. Lopez, and \DVN, \TAMU{61}/92.}
\nref \HMY{J. Hisano, H. Murayama, and T. Yanagida, Tohoku University preprint
TU-400/92.}
\nref\ZUT{H. Dreiner, J.L. Lopez, D.V. Nanopoulos and
D. Reiss, \PLB{216}{89}{289}.}
\nref\FLIP{I. Antoniadis, J. Ellis, J. Hagelin, and D.V. Nanopoulos,
\PLB{194}{87}{231};
{\bf B205} (1988) 459;  {\bf B208} (1988) 209;
{\bf B231} (1989) 65.}
\nref\SSM{L. Ibanez and G. G. Ross, CERN preprint CERN-TH.6412/91.}
\nref\SPAR{A. Faragii, J. Hagelin, S. Kelley, and \DVN, \PRD{45}{92}{3272}.}
\nref\KAP{J. Minahan, \NPB{298}{88}{36};\nextline
V. Kaplunovsky, \NPB{307}{88}{145}.}
\nref\MOD{L.E. Ibanez, D. Lust and G.G. Ross, CERN preprint CERN-TH.6241/91}
\nref\SISM{S. Kelley, J. Lopez, and \DVN, PLB{278}{92}{140}.}
\nref\EF{J. Ellis, G. Fogli and E. Lisi, \PLB{274}{92}{456} and
references therein.}
\nref\DEGRASSI{G. Degrassi, S. Fanchiotti and A. Sirlin,
\NPB{351}{91}{49}.}
\nref\HADRONIC{H. Burkhardt, F. Jegerlehner, G. Penso and
C. Verzegnassi,
Z. Phys. {\bf{C11}} (1989) 497.}
\nref\ASTRONG{ALEPH Collaboration, D. Decamp et al.,
CERN Preprint PPE/92-33 (1992); \nextline
DELPHI Collaboration, P. Abreu et al., CERN Preprint
PPE/91-181/Rev. (1992);\nextline
L3 Collaboration, O. Adriani et al., CERN Preprint
PPE/92-58 (1992);\nextline
OPAL Collaboration, P.D. Acton et al., CERN Preprint
PPE/92-18 (1992);\nextline
A.X. El Khadra, G. Hockney, A.S. Kronfeld and
P.B. Mackenzie, Fermilab Preprint PUB-91/354-T (1991);\nextline
W. Kwong, P.B. Mackenzie, R. Rosenfeld and J.L. Rosner,
\PRD{37}{88}{3210};\nextline
A.D. Martin, R.G. Roberts and W.J. Stirling, Durham Preprint
DTP 90-76 (1990), Rutherford Preprint RAL-91-044 (1991).}
\nref\ROSS{J. Ellis, D.V. Nanopoulos and D. Ross, \PLB{267}{91}{132}.}
\nref\WINDOW{I. Antoniadis, J. Ellis, and \DVN, \PLB{262}{91}{109}.}
\nref\PLIMIT{K. Hikasa {\it et al.} (Particle Data Group), PRD{45}{92}{1}.}
\nref\EKNIII{J. Ellis, S. Kelley and \DVN, CERN preprint CERN-TH.6140/91
(1991), to be published in Nucl. Phys. B.}
\nref\BARB{R. Barbieri, L.J. Hall, LBL preprint LBL-31238/91}
\nref\ACPZ{F. Anselmo, L. Cifarelli, A. Petermann and A. Zichichi,
CERN-TH.6429/92 (1992).}
\nref\ZIC{F. Anselmo, L. Cifarelli, A. Petermann and A. Zichichi,\nextline
CERN/LAA/MSL/91-015; CERN/LAA/MSL/91-026; CERN/LAA/MSL/92-008; \nextline
CERN/LAA/MSL/92-009; CERN/LAA/MSL/92-011; CERN/LAA/MSL/92-012.}
\nref\EWX{S. Kelley, J. Lopez, \DVN, H. Pois, and K. Yuan, CERN-TH.6498/92.}
\nref\NAT{J. Ellis, K. Enqvist, \DVN, and F. Zwirner,
\MODA{1}{86}{57};\nextline
R. Barbieri, and G. Giudice, \NPB{306}{88}{63}.}

\nfig\I{Figure 1 shows constraints in the $\st$, $\as$ plane from proton decay
and coupling constant unification without threshold effects.
The two ellipses give $1-\sigma$
experimental areas for the average $\as=.122$ of the first five high values
of $\as$ and the average $\as=.109$ of the
last three low values of $\as$ given in Table 2.  The dark line shows
the prediction for $\st$ from supersymmetric $SU(5)$ without threshold effects.
The width of the line corresponds to the uncertainty in $\ae$.  This
dark line is an upper bound on $\st$ in the Flipped model.  The two dotted
lines represent a lower bound on $\st$ in the Flipped model from proton decay,
with the top dotted line using a hadronic matrix element of
$\alpha=.03$, and the bottom dotted line using a hadronic matrix element of
$\alpha=.003$.  The two lighter solid lines represent the prediction from
non-supersymmetric $SU(5)$, with the bottom line corresponding to a one higgs
doublet model, and the top line corresponding to a two higgs doublet model.}

\nfig\II{The prediction for the average supersymmetric mass scale as a function
of $\as$ assuming $\delta_s(heavy)=0$ is shown as the area between the two
lines.
At least one field must have mass in this range.  For $\delta_s(heavy)>0$, only
the
lower bound remains.  To ensure that no field has mass above the $1\TeV$ dashed
line,
$\as$ must be above about 0.11.}

\nfig\III{An explicit parameterization of the light thresholds neglecting the
EGM effect excludes the parameter space to the left of the solid line which
gives
the prediction for $m_{1/2}$ as a function of $\as$ for the $X_{max}$ given in
Table 4.
A reasonable value
for the soft supersymmetry-breaking gaugino mass represented by the dashed
lines,
$45\GeV<m_{1/2}<1\TeV$, and a $1-\sigma$
range of the other inputs constrains the strong coupling, $\as>.114$.}

\nfig\IV{Same as Figure 3, but including the EGM effect.  The main effect
is to modify the slope of the boundary for high values of $m_{1/2}$
loosening the bound in those regions.  Bounds for other values of $X$
summarized
in Table 4 are shown as dashed and dotted lines for reference.}

\nfig\V{One-loop boundaries for the SSM with all consistency and
phenomenological
cuts imposed for $m_{1/2}=150,250\GeV$, both signs of $\mu$ and the following
($\xi_0,\xi_A$) values: (a) (0,-1)(dashed), (0,0)(solid), (0,1)(dotted);
(b) (1,-1)(dashed), (1,0)(solid), (1,1)(dotted).  The figures show the
progression
of the left boundary to higher values of $m_t$ due to the unit variations of
$\xi_0$ and $\xi_A$.}

\Titleh{\vbox{\baselineskip12pt
\hbox{MIU--THP--92/61}
\hbox{October, 1992}}}
{\vbox{\centerline{
Evidence For SUSY From GUTS?}
\vskip2pt\centerline{
Evidence For GUTS From SUSY!}}}
\centerline{John S. Hagelin and S. Kelley}
\medskip
\centerline{{\it Department of Physics}}
\centerline{{\it Maharishi International University}}
\centerline{{\it Fairfield, Iowa 52557-1069 USA}}
\bigskip
\vskip .2in
\font\abstractfont=cmbx12
\long\def\abstract#1//{%
  \centerline{\abstractfont Abstract}
  \vskip .1in
  {\baselineskip=14pt\advance\leftskip by 2pc\advance\rightskip by
2pc\parindent=10pt
  \def\enspace{\kern.3em}
  \noindent #1\par}}
\abstract We review the theoretical and experimental status of minimal grand
unified theories (GUTS), contrasting the failure of \mnsg\ with the success of
the \msg\ and \msfg\ models.  We show that a reasonable value for the universal
soft supersymmetry-breaking gaugino mass,
$45\GeV<m_{1/2}<1\TeV$, and a $1-\sigma$
range of the other inputs constrains the strong coupling, $\as>.114$.  We
define the
supersymmetric standard model (SSM), the minimal supersymmetric extension of
the
standard model with gauge coupling unification and universal soft
supersymmetry-breaking
at the unification scale, as a baseline model for unified theories.  We review
the
structure of the allowed parameter space of the SSM and suggest
sparticle spectroscopy as the
experimental means to determine the parameters of the SSM and search for
departures from the baseline SSM.//
\vskip .4in
\centerline{Invited talk at the Ten Years of SUSY Confronting Experiment
conference}
\centerline{CERN, September 7--9, 1992}
\bigskip
\vskip .2in
\bgroup\baselineskip=12pt
\leftline{MIU--THP--92/61}
\leftline{October, 1992}
\egroup
\Date{}
\newsec{Overview}

In addition to the major theoretical problem of the gauge hierarchy, minimal
non-supersymmetric
$SU(5)$ \GeorgiI\
flunks the test of proton decay and $\st$.
Although supersymmetry technically solves the gauge hierarchy problem,
a fine tuning problem re-emerges in \msg\ \SU\ in splitting
the triplet and doublet components of the five dimensional matter
superfields in the model: the triplets must have GUT scale masses, while
the doublets must remain light to yield electroweak scale symmetry
breaking.

The prediction of \msg\ for $\st$ matches so closely the
experimental value \EKNI\ that one may hope to constrain the spectum of the
model through the threshold contributions to this prediction \AMALDI\EKNII.
This
shall be the main focus of this work.

Supersymmetrizing the minimal $SU(5)$ model increases the GUT scale, thus
ensuring that the dimension six proton
decay operators, which doomed the minimal non-supersymmetric theory, are no
problem
in the supersymmetric theory.  However, dimension five operators give proton
decay
very near the experimental limit and sensitive to the details
of the supersymmetric spectrum.  This allows constraints to be placed
on the supersymmetric spectum of the model from the non-observation of
proton decay \PDECAY.  One study shows that the combined constraints of proton
decay, naturalness, and a neutralino relic density smaller than the
closure density severely constrains the model \LPN.  However, another study
finds that experimental values of the low-energy couplings allow a larger
higgs triplet mass than had been previously considered, which considerably
relaxes the constraints from proton decay \HMY.  Similarly, as will be
emphasized,
the constraints from coupling constant unification are extremely sensitive
to the experimental inputs and subtleties of the actual calculation.
The two questions of proton decay and coupling constant are interdependent as
both the light and heavy spectum of the model
enter each prediction.

Besides the doublet-triplet splitting problem of \msg,
the model faces another challenge when one tries to reconcile it
with the only available consistent theory of quantum gravity, the string.
The adjoint representation needed to break the $SU(5)$ symmetry is unavailable
in string theories at Kac-Moody level k=1\ZUT.
Though someday, realistic string models with higher
Kac-Moody levels may be possible, there is a simpler and more elegant GUT
whose particle spectrum is available in the string at level $k=1$.  This \msfg\
GUT \FLIP\ has a natural doublet-triplet splitting
mechanism which eliminates the fine-tuning problem and reduces the effect of
the dimension five proton-decay operators well below that of the dimension
six operators.  The prediction for $\st$ for supersymmetric $SU(5)$
becomes an upper bound for the prediction of $\st$ in \msfg\ because
of the extra scale.  Thus, the constraints
from proton decay and coupling constant unification in the Flipped model
are considerably less stringent than in the \msg\ model.

Although, coupling constant unification and proton decay rule
out the minimal non-supersymmetric $SU(5)$ GUT,this only constitutes
circumstantial evidence for supersymmetry.
All of the problems of \mnsg\ could probably be fixed,
except perhaps the hierarchy problem,
by appropriate non-supersymmetric extensions of the minimal model.
The real verification of
supersymmetry will be direct observation of spartners.

To simplify extracting
low-energy predictions of unified models, it is useful to eliminate the
model-dependent GUT structure and consider a minimal supersymmetric
extension of the standard model with coupling constant unification and
universal soft supersymmetry breaking at the unification scale.  We refer to
this
model as the supersymmetric standard model (SSM), which has been
extensively studied (for a recent review see \SSM).  The
low energy-predictions of the SSM are very near those of supersymmetric
$SU(5)$ and supersymmetric Flipped $SU(5) \times U(1)$ and can be used as a
baseline to
search for departures indicating a particular unified theory.  The
SSM has five unmeasured parameters, the top mass $m_t$, the ratio of higgs vevs
$tan\beta$,
and the three soft susy-breaking parameters $m_{1/2}$, $m_0$, and $A$.
The entire spectrum and S-matrix
of this model can be calculated for any point in this five-dimensional
parameter space.  Electroweak breaking and experimental constraints then
give a boundary between allowed and disallowed points in this five dimensional
parameter space.

The spectum of sparticles corresponding to the two light generations of
fermions has a particularly simple structure which depends only on
$tan\beta$, $m_{1/2}$, and $m_0$.
Measurement of three sparticle masses determines the values of $tan\beta$,
$m_{1/2}$,
and $m_0$
with a fractional uncertainty for $m_{1/2}$ and $m_0$
comparable to that of the mass measurements \SPAR.
The discussion in the SDC talk at this conference
of a 10\% resolution for the gluino mass gives
a first indication that this program of sparticle spectroscopy may be
feasible.  Sooner or later, we must face the necessity of verifying GUT
scale physics with precision low-energy experiments.  In addition to looking
for
proton-decay, lepton flavour violation, and other rare decays, the sparticle
spectrum's
sensitivity to almost every detail of a theory makes it an ideal place to
look for evidence for GUTS from SUSY.

The unification scale of the class of minimal supersymmetric theories like
$SU(5)$ and $SU(5) \times U(1)$, whose low-energy predictions are close to
that of the SSM, turns out to be about $10^{16}\GeV$.  The string gives a
coupling constant unification scale of about $10^{18}\GeV$ \KAP.  One way of
reconciling the discrepancy between the unification scale predicted by the
string and
that predicted by the SSM
is to look for string models where moduli dependent corrections
bring the string unification scale down to $10^{16}\GeV$ \MOD.
Alternatively, adding massive vector representations to the SSM can bring its
unification
scale up to $10^{18}\GeV$.  The minimal set of massive representations with
standard-model-like quantum numbers which could accomplish this are a vector
pair of quark
doublets with mass of about $10^{13}\GeV$ and a vector pair of right-handed
down quarks with mass of about $10^5\GeV$.  This model has been named the
String Inspired Standard Model (SISM) \SISM.  Varying the mass of the two
vector pairs
allows $\st$ and the unification scale to match respectively the experimental
value
and the string unification scale.  The SISM
represents the minimal particle content needed to do so and many other
extra-vector models exist with more than this minimal content.  The
SISM depends on the same five parameters as the SSM and the SISM low energy
predictions have been found to be qualitatively similar but quantitatively
different than the predictions of the SSM \SISM.

Table 1. summarizes the various tests of the models discussed in this section.

\newsec{What Is The Strong Coupling?}

Coupling constant unification and the GUT scale depend on the value of
the low energy couplings.  All these inputs will be taken at $m_Z$ in the
$\overline{MS}$ renormalization scheme.  The $1-\sigma$ values of the
electromagnetic
coupling and $\st$ we use are:

\eqn\I{\st=0.2328\pm 0.0009 \quad \EF}

\eqn\II{\ae={1\over{127.9\pm 0.2}}\quad  \DEGRASSI\HADRONIC.}

However, a glance at Table 2, which summarizes different determinations
of the strong coupling, \ASTRONG\ indicates a problem with specifying the value
of $\as$: although the individual measurements have fairly small errors,
different
determinations of the strong coupling give very different results.  These
results fall into two main classes: the first five LEP measurements at $m_Z$
which average to $\as=.122$, and the last three low energy measurements
extrapolated to $m_Z$ which average to $\as=.109$.  One suggestion
is that higher order QCD corrections to jet shapes reduce the high LEP
measurements \ROSS.  Another suggestion is that the gluino mass is in the
swiftly shrinking light gluino window.  If this were the case, the QCD beta
function would include gluino contributions when running the low
energy measurements up to $m_Z$.  Doing this brings the low-energy measurements
of $\as$ into
amazing agreement with the LEP measurements \WINDOW.  Whatever the details of
the
explanation, it seems certain that the proper inclusion of the different
radiative corrections to each type of measurement of the strong coupling is the
key
to resolving this problem.  Because of this uncertainty, the results of
this talk will be presented as bounds on the strong coupling.

\newsec{Dimension Six Proton Decay}

Table 3 summarizes the bounds from dimension six proton decay
on the mass of the superheavy gauge bosons and the strong coupling using a
limit on the partial lifetime of the proton
\eqn\III{\tau(p\rightarrow e^+\pi^0)<5.5\times 10^{32}yr\quad\PLIMIT.}
The two values in Table 3 for each model correspond to the extremes of an
order-of-magnitude
uncertainty in the hadronic matrix element $.003<\alpha<.03$.  The minimal
non-supersymmetric models are ruled out, while the minimal supersymmetric
models have no trouble with dimension six proton decay.  Note however that
the parameter space of minimal supersymmetric $SU(5)$ is severely constrained
by dimension five proton decay \PDECAY. In supersymmetric Flipped $SU(5)\times
U(1)$,
the unification scale depends on both $\as$ and $\st$ so the results
must be presented in the $\st$, $\as$ plane.

\newsec{Constraints in the $\st$, $\as$ Plane Without Thresholds}

Figure 1 shows constraints in the $\st$, $\as$ plane from proton decay
and coupling constant unification without threshold effects.
The two ellipses give $1-\sigma$
experimental areas for the average $\as=.122$ of the first five high values
of $\as$ and the average $\as=.109$ of the
last three low values of $\as$ in Table 2.  To be conservative,
the errors of the average high and low $\as$ ellipses have been taken
as the smallest error of the individual experiments, $\pm .005$.  The dark line
shows
the prediction for $\st$ from supersymmetric $SU(5)$ without threshold effects.
The width of the line corresponds to the uncertainty in the $\ae$.  This
solid line is an upper bound on $\st$ in the Flipped model.  The two dotted
lines represent a lower bound on $\st$ in the Flipped model from proton decay,
with the top dotted line using a hadronic matrix element of
$\alpha=.03$, and the bottom dotted line using a hadronic matrix element of
$\alpha=.003$.  The two lighter solid lines represent the prediction for $\st$
from
non-supersymmetric $SU(5)$, with the bottom line corresponding to a one higgs
doublet model, and the top line corresponding to a two higgs doublet model.
Both
minimal non-supersymmetric $SU(5)$
models are many $\sigma$ off in their prediction for $\st$, while the
two supersymmetric models are right on the money.

\newsec{Threshold Corrections in the Minimal Supersymmetric $SU(5)$ GUT}

The prediction for $\st$ in minimal supersymmetric $SU(5)$ may be
written as \EKNIII
\eqn\IV{\st=0.2+{{7\ae}\over{15\as}}+0.0029+
\delta_s(light)+\delta_s(heavy)+\delta_s(conv)}
where 0.0029 corrects the analytic one-loop calculation to two-loop
accuracy, $\delta_s(light)$ gives the correction from light thresholds, and
$\delta_s(heavy)$ gives the correction from heavy thresholds.
The scheme conversion term $\delta_s(conv)$ is negligible.

If, for a moment, we assume $\delta_s(heavy)=0$ and simply parameterize the
light
fields by $m_t$, half the higgs degrees of freedom at or below $m_Z$, and the
rest
of the fields
beyond the Standard Model in the SSM degenerate at a scale $m_{SUSY}$,
$\delta_s(light)$ becomes
\eqn\V{\delta_s(light)={{\ae}\over{20\pi}}\biggl[
-3\ln\Bigl({{m_t}\over{m_Z}}\Bigr)
-{{19}\over 3}\ln\Bigl({{m_{SUSY}}\over{m_Z}}\Bigr)\biggr],}
and \IV\ can be solved for $m_{SUSY}$ in terms of the other inputs.

Using the range of $\st$ and $\ae$ given in \I\ and \II,
and  a $1-\sigma$ global fit, $92\GeV<m_t<147\GeV$, from $\st$ \EF\
gives a range of allowed values for $m_{SUSY}$ as
a function of $\as$ shown as the band between the two lines in Figure 2.
Since the spectrum is not actually degenerate, what this really
means is that there must be at least one field with mass in this range!
This gives the remarkable conclusion that for $\as>.118$,
there must be at least one new field with mass less than about $1\TeV$.

With a general GUT structure, the sign and magnitude of $\delta_s(heavy)$
is uncertain, and washes out this conclusion \BARB.  But, the minimal
supersymmetric $SU(5)$ GUT has a very simple heavy threshold contribution:
\eqn\VI{\delta_s(heavy)={{\ae}\over{20\pi}}\biggl[
-6\ln\Bigl({{M_{GUT}}\over{M_{D^C}}}\Bigr)
+4\ln\Bigl({{M_{GUT}}\over{M_{V}}}\Bigr)+2\ln\Bigl({{M_{GUT}}\over{M_{\Sigma}}}\Bigr)\biggr],}
where $M_{D^c},M_V,M_{\Sigma}$ are the masses of the proton-decay mediating
higgs
triplet matter superfields, super-heavy gauge superfields, the uneaten remnants
of
the $SU(5)$ adjoint matter superfield, and
$M_{GUT}=max(M_{D^c},M_V,M_{\Sigma})$ is
the scale where the couplings become equal.
Note that the only possibility of a negative contribution
from $\delta_s(heavy)$ is if $M_{D_c}<M_{GUT}$.  Since for reasonable values of
$\as$,
$M_{GUT}$ cannot be much greater than $10^{16}\GeV$,
and $M_{D_c}=10^{16}\GeV$ already gives substantial constraints on
the SUSY spectrum, it is likely that $M_{D_c}=M_{GUT}$, especially for small
$\as$ where
$M_{GUT}$ is even lower.  Since the regions that will be eventually constrained
are $\as<.114$, it is safe to take  $\delta_s(heavy)>0$.
To effectively explore the possibility that $M_{D_c}<M_{GUT}$
would require an analysis of the combined constraints on the supersymmetric
spectrum from both proton decay and coupling constant unification.

The requirement that $\delta_s(heavy)>0$ leaves only the lower bound
on $m_{SUSY}$ which translates into at least one field with mass
above the lower line in Figure 2.  For $\as$ less than about $0.11$, the graph
shows
that there must
be at least one field with mass greater than $1\TeV$.
If one's naturalness criteria forbids supersymmetric fields greater than
$1\TeV$, then $\as<.11$ is excluded in the \msg.
This type of analysis \AMALDI\ reveals the essential physics, and the following
sections derive even tighter bounds on $\as$ in the \msg\ model by using a more
detailed
parameterization of the light thresholds.

\newsec{An Explicit Parameterization of the Light Thresholds}

The contribution to $\st$ from light particle thresholds was derived in
\EKNIII:
\eqna\VII
$$\eqalignno{\delta_s(light)=&{{\ae}\over{20\pi}}
\biggl[-3\ln\Bigl({{m_t}\over{m_Z}}\Bigr)
+{28\over 3}\ln\Bigl({c_{\tilde g}m_{1/2}\over m_Z}\Bigr)
-{32\over 3}\ln\Bigl({c_{\tilde w}m_{1/2}\over m_Z}\Bigr)\cr
&-\ln\Bigl({{m_h}\over{m_Z}}\Bigr)-4\ln\Bigl({{\mu}\over{m_Z}}\Bigr)+{4\over
3}f(y,w)\biggr]
&\VII {}\cr}$$
where
\eqna\VIII
$$\eqalignno{
f(y,w)=&{{15}\over 8}\ln\bigl({{\sqrt{c_{\tilde q}+y}}}\hskip2pt\bigr)
-{9\over 4}\ln\bigl({{\sqrt{c_{\tilde e_l}+y}}}\hskip2pt\bigr)\cr
&+{3\over 2}\ln\bigl({{\sqrt{c_{\tilde e_r}+y}}}\hskip2pt\bigr)
-{{19}\over{48}}\ln\bigl({{\sqrt{c_{\tilde q}+y+w}}}\hskip2pt\bigr)\cr
&-{{35}\over{48}}\ln\bigl({{\sqrt{c_{\tilde q}+y-w}}}\hskip2pt\bigr)
&\VIII {}\cr}$$
and $y \equiv (m_{0}/m_{1/2})^2$, $m_{0}$ is a universal primordial
supersymmetry-breaking spin-zero mass, $w$ was defined in \EKNIII\
and the logarithms should be set to zero if the threshold is below $m_Z$.

With this parameterization for $\delta_s(light)$ and taking
$\delta_s(heavy)>0$,
in the region where $m_{\tilde w}<m_Z<m_{\tilde g}$, \IV\ can be manipulated to
yield
\eqn\IX{
\ln\Bigl({{m_{1/2}}\over{m_Z}}\Bigr)<{1\over 7}X
-{\pi\over{\alpha_3}}-\ln(c_{\tilde g})}
where
\eqn\X{
X={{15\pi}\over{\ae}}(\st-.2029)
+{9\over 4}\ln\Bigl({{m_t}\over{m_Z}}\Bigr)+3\ln\Bigl({{\mu}\over{m_Z}}\Bigr)
+{3\over 4}\ln\Bigl({{m_h}\over{m_Z}}\Bigr)
-f(y,w)}
Now consider the region where both the gluinos and the winos are heavier
than $m_Z$, in which case \IV\  gives:
\eqn\XI{
\ln\Bigl({{m_{1/2}}\over{m_Z}}\Bigr)>
-X+{{7\pi}\over{\alpha_3}}+7\ln(c_{\tilde g})-8\ln(c_{\tilde w})}
The most generous bounds from \IX\  and \XI, result from
maximizing $X$, minimizing $c_{\tilde g}$, and
maximizing $c_{\tilde w}$.

For physically-relevant values of $w$ (those which give positive squared masses
for the stop squarks), $f(y,w)$
is minimized at $w=-8(c_{\tilde q}+y)/27$.
With this value of $w$, $f(y,w)$ has one extremum, a maximum, at
$y=(c_{\tilde l_l}c_{\tilde l_r}+2c_{\tilde l_l}c_{\tilde q}
-c_{\tilde l_r}c_{\tilde q})/(3c_{\tilde l_l}-2c_{\tilde l_r}-c_{\tilde q})$,
and approaches -0.025 as y becomes very large.  Since the values of the
$c's$ that we encounter satisfy $c_{\tilde q}>1>c_{\tilde l_l},c_{\tilde l_r}$,
the minimum of $f(y,w)$ is indeed -0.025 for values of $y>0$.
We have numerically searched the region where the
scalars are lighter than $m_Z$ to verify
that $f(y,w)$ does not take smaller values in this region.
This minimum value of $f(y,w)=-0.025$ maximizes $X$.

Taking the extreme $1-\sigma$ values $\st=0.2337$, $m_t=92\GeV$,
$\ae=1/128.1$ and using a
naturalness bound of $1\TeV$ for $\mu$ and $m_h$ gives a
maximum numerical value of $X=195.0$.  Note that the
extreme values of $\st$ and $m_t$ are correlated, and that the contribution
of the squarks and sleptons represented by $f(y,w)$ is
negligible.  Table 4 shows the sensitivity of $X$ to the inputs: central values
for $\st$, $m_t$, and $\ae$ give $X=189.8$, and extreme $1-\sigma$ values of
the inputs give a minimum $X=184.6$.  However, because minimum values of
$\delta_s(heavy)$ and $f(y,w)$ have been used to maximize $X$, only the upper
bound
$X<195.0$ really matters.

The ratios of the gaugino masses to the universal soft supersymmetry-breaking
gaugino mass are
\eqn\XII{c_{\tilde g}={{\alpha_3(m_{\tilde g})}\over{\alpha(M_{GUT})}} \quad
c_{\tilde w}={{\alpha_2(m_{\tilde w})}\over{\alpha(M_{GUT})}}.}
Approximating $c_{\tilde g}$ and $c_{\tilde w}$ at $m_Z$ without including
threshold effects gives
\eqn\XIII{c_{\tilde g}\approx 2.7 \quad c_{\tilde w}\approx 0.79,}
which turns out to be a very bad assumption \ACPZ.

Putting all this together gives a bound on $m_{1/2}$ as a function of $\as$
excluding the parameter space in the \msg\ to the left of the line in Figure 3.
To have a reasonable soft supersymmetry breaking scale $45\GeV<m_{1/2}<1\TeV$,
the strong coupling is bounded by $\as>.114$.  The next section
shows how a correct treatment of the gaugino masses effects this bound.

\newsec{The EGM Effect}

Since the gaugino masses should be computed using $c_{\tilde g}$ and
$c_{\tilde w}$ evaluated at the gaugino mass, the numerical values used for
the $c's$ in the previous section \XIII\ become increasingly inaccurate for
higher gaugino masses.  This evolution of the gaugino mass (EGM) effect \ACPZ\
and
several other subtle points in the computation of gauge coupling unification
have been extensively studied \ZIC.  Since the gaugino masses where the $c's$
should be evaluated depends on the value of $m_{1/2}$, \IX\ and \XI\ must
be evaluated iteratively, recomputing the $c's$ at each iteration.
This can be simplified by realizing the bounds on $m_{1/2}$ remain rigorous
using
a minimum for $c_{\tilde g}$ and a maximum for $c_{\tilde w}$.

 From the one-loop expression for renormalizing a coupling
from $m_Z$ up to its gaugino threshold,
\eqn\XIV{
\alpha(m_{gaugino})={{\alpha(m_Z)}\over{1-{b\over{2\pi}}
\ln({{\alpha(m_{gaugino})m_{1/2}}\over{m_Z\alpha_U}})}},}
we see that, for $b<0$, $\alpha(m_{gaugino})$
increases with $b$.  In order to minimize $c_{\tilde g}$, we want
to use the minimum value of $b_3=-7$ possible in the SSM below the
gluino threshold.  Similarly, to maximize
$c_{\tilde w}$ we use the maximum value of $b_2=-1/3$ possible
below the wino threshold.

Fitting the results of an analytic
one-loop calculation to a numeric two-loop
calculations for central values gives \EKNIII:
\eqna\XV
$$\eqalignno{{1\over{\alpha_U}}=
&{3\over{20\ae}}+{3\over{5\alpha_3(m_Z)}}-0.7
+{1\over{5\pi}}\biggl[3\ln\Bigl({{m_{\tilde g}}\over{m_Z}}\Bigr)
+{1\over 8}\ln\Bigl({{m_{h}}\over{m_Z}}\Bigr)
+{3\over 8}\ln\Bigl({{m_{\tilde l_l}}\over{m_Z}}\Bigr)\cr
&+{3\over 8}\ln\Bigl({{m_{\tilde l_r}}\over{m_Z}}\Bigr)
+{{17\over 4}}\ln\Bigl({{m_{\tilde q}}\over{m_Z}}\Bigr)
+{{83}\over{48}}\ln\Bigl({{m_t}\over{m_Z}}\Bigr)
+{1\over 2}\ln\Bigl({{m_{\tilde w}}\over{m_Z}}\Bigr)
+{1\over 2}\ln\Bigl({{\mu}\over{m_Z}}\Bigr)\biggr]&\XV {}\cr}$$
where the stop squarks have been taken degenerate with the other squarks.
Thus, $\alpha_U$ decreases with the thresholds.  Taking upper bounds of
$147\GeV$ on the top mass and $3\TeV$ on the other thresholds gives the range
\eqn\XVI{{3\over{20\alpha_{em}}}+{3\over{5\alpha_3(m_Z)}}-0.7<
{1\over{\alpha_U}}<
{3\over{20\alpha_{em}}}+{3\over{5\alpha_3(m_Z)}}+1.4}
for the coupling at the unification scale.
Numerically, we find a slight variation of the
solutions of \IX\ and \XI\ over
this range of $\alpha_U$, with both values increasing with $\alpha_U$.
Therefore, we use the maximum value in \IX\ and the minimum value in \XI.

The results of this calculation are shown in Figure 4.  The region to the left
of the solid line is excluded in the \msg.  Bounds for $X_{central}$ and
$X_{max}$
are shown as dashed and dotted lines for reference.  To
have a reasonable range for the universal soft supersymmetry-breaking
gaugino mass, $45\GeV<m_{1/2}<1\TeV$, between the horizontal dashed lines in
Figure 4,
the strong coupling is constrained by $\as>.114$.  Note that
the EGM effect modifies the slope of the bound for $m_{\tilde w}>m_Z$.
However, this has little effect on the overall bound for $\as$ which comes
from low $m_{1/2}$ regions where the EGM effect is small.

\newsec{The Parameter Space of the SSM and Sparticle Spectroscopy}

By cleverly using the constraints from electroweak symmetry breaking, the SSM
can be described by five unknown parameters; $m_t$, $tan\beta$, $m_{1/2}$,
$m_0$, and $A$ \EWX.  The last two parameters are more conveniently expressed
in terms of the dimensionless parameters $\xi_0=m_0/m_{1/2}$ and
$\xi_A=A/m_{1/2}$.
The parameter space splits into two sections, one for
each sign of the higgs mixing parameter $\mu$, its magnitude determined by
radiative breaking in terms of the other parameters.
The light quark masses, KM elements, and gauge couplings, despite
experimental errors, are considered as known.  Furthermore, only the bottom and
tau mass
and the gauge couplings effect the supersymmetric spectrum significantly.  This
amazing simplification over a generic supersymmetric extension of the Standard
Model
results from the assumptions of universal soft supersymmetry-breaking at the
unification scale in the SSM.  Because the whole supersymmetric spectrum
results
from just five parameters, of which $m_t$ should soon be known, the model is
extremely predictive.

For the yukawa couplings to remain perturbative up to the unification scale,
$m_t$ must be less than about $190\GeV$ and $tan\beta$ less than about $50$.
Since radiative breaking requires $tan\beta>1$ and experiment gives
$m_t>90\GeV$,
the parameter space is completely bounded in $m_t$ and $tan\beta$.  The values
of
the soft supersymmetry-breaking parameters can be bounded from above by
naturalness \NAT, but the exact bounds remain somewhat a matter of taste.

To get a feeling for this parameter space, Figure 5 shows some
representative slices \EWX.  In these figures, the solid line corresponds
to $\xi_A=0$, the dotted lines to $\xi_A=1$, and the dashed lines to
$\xi_A=-1$.
Computer visualization can be used to show a three-dimensional slice of the
parameter space, and even a four-dimensional slice as a movie.  Some first
attempts in this direction were seen at this conference.

The allowed parameter space of the SSM is huge.  However, knowing a few
sparticle masses would very quickly narrow it.  The sparticles corresponding to
the two light generations have a very simple dependence on only three of the
SSM parameters.

\eqn\XVII{m^2_{\tilde p}=m_0^2+c_{\tilde p}(m_{\tilde p})m^2_{1/2}+
2\Bigl[T^{\tilde p}_{3_L}-{3\over 5}Y^{\tilde p}tan^2\theta_W\Bigr]m^2_W}

\noindent Measurements of three sparticle masses can be translated into a
determination of
$tan\beta$, $m_{1/2}$, and $m_0$ with fractional uncertainties in the
determination of $m_{1/2}$, and $m_0$ comparable to the fractional
uncertainties
of the sparticle masses \SPAR.

Sufficiently accurate determination of more sparticle masses could be used to
discriminate between different extensions of the SSM such
as extensions of the Standard Model gauge group, additional Yukawas,
generational-dependent extra heavy gauge bosons, and non-universal
supersymmetry-breaking which all leave distinct imprints on the sparticle
spectrum.

\newsec{Conclusions}

The success of minimal supersymmetric GUTS compared to the failure of
minimal non-supersymmetric GUTS gives strong circumstantial evidence that a
viable GUT should be supersymmetric.  The interplay between naturalness,
proton decay, and coupling constant unification provides a tool to constrain
the parameter space of supersymmetric GUTS.  These constraints begin to rule
out areas in the minimal supersymmetric $SU(5)$ model.  In particular, to
have a reasonable range for the universal soft supersymmetry-breaking
gaugino mass $45\GeV<m_{1/2}<1\TeV$,
the strong coupling is constrained by $\as>.114$.

The SSM, the minimal supersymmetric extension of the Standard Model with
coupling constant unification and universal soft supersymmetry-breaking
at the unification scale, closely reproduces the low-energy structure
of supersymmetric GUTS.  Five unknown parameters $m_t$, $tan\beta$, $m_{1/2}$,
$m_0$, and $A$ specify the entire spectrum and S-matrix of the model.
Electroweak symmetry breaking and experimental constraints can be used to
produce a boundary between allowed and disallowed regions of parmeter space
in the model.

Measuring the masses of three sparticles corresponding to the two light
generations could
be used to experimentally extract the SSM values of $tan\beta$, $m_{1/2}$, and
$m_0$ with
fractional uncertainty of $m_{1/2}$, and $m_0$ comparable to that of the mass
determinations of the sparticles.  Other spartner masses could be use to
determine $A$
and check the consistency of the SSM for corrections due to a GUT structure
which
leave a distinct imprint on the sparticle spectrum regardless of the scale of
extra physics.

Now that GUTS have revealed the need for SUSY, it is time to test GUTS by
measuring
SUSY!

\vskip 0.5cm
\noindent
{\bf Acknowledgements}

The work of J.S.H. was supported in part by the National Science Foundation
under
grant no. PHY-9118320.
One of us (S.K.) thanks J. Ellis and \DVN\ for their support and inspiration.

\vskip 0.5cm

\input tables
\vfill \eject
\vskip 2cm
{\hfill
{\begintable
\ Test \ \|\ Non-Susy $SU(5)$ \ \|\ Susy $SU(5)$ \ \|\ Susy $SU(5)\times U(1)$
\|\ SSM \|\ SISM
 \crthick
$\st$
\|\ X \| ? \| $\surd$ \| $?$ \| $\surd$ \nr
proton decay
\|\ X \| ? \| $\surd$ \| $\surd$ \| $\surd$  \nr
$m_b/m_{\tau}$
\|\ $\surd$ \| $\surd$ \| $\surd$ \| $\surd$ \| $\surd$  \nr
fine-tuning
\|\ X \| $\surd$ \| $\surd$ \| $\surd$ \| $\surd$  \nr
doublet-triplet splitting
\|\ - \| X \| $\surd$ \| - \| - \nr
k=1 string
\|\ X \| X \| $\surd$ \| $\surd$ \| $\surd$  \nr
$M_U$
\|\ $\approx 10^{14}$ \| $\approx 10^{16}$ \| $\approx 10^{16}$ \| $\approx
10^{16}$ \| $\approx 10^{18}$  \endtable}

\hfill}
\smallskip
\parindent=0pt
\centerline{
{\it Table 1} - Comparison of minimal unified models.}

\vfill \eject
\vskip 2cm
{\hfill
{\begintable
\ Experiment \ \|\ Central Value \ \|\ Error \crthick
ALEPH  jets \|\ ~~$0.125$ \| ~~$\pm 0.005$ \nr
DELPHI jets \|\ ~~$0.113$ \| ~~$\pm 0.007$ \nr
L3 jets \|\     ~~$0.125$ \| ~~$\pm 0.009$\nr
OPAL   jets \|\ ~~$0.122$ \| ~~$\pm 0.006$ \nr
OPAL   $\tau$ \|\ ~~$0.123$ \| ~~$\pm 0.007$ \nr
$J/\Psi$ \|\ ~~$0.108$ \| ~~$\pm 0.005$ \nr
$\Upsilon$ \|\ ~~$0.109$ \| ~~$\pm 0.005$ \nr
Deep Inelastic \|\ ~~$0.109$ \| ~~$\pm 0.005$ \nr
Average \|\ ~~$0.116$ \| ~~$\pm 0.005$ \endtable}
\hfill}
\smallskip
\parindent=0pt
\centerline{
{\it Table 2} - Experimental Values of $\alpha_3(m_Z)$.}

\vskip 2cm
\vfill
\break
{\hfill
{\begintable
\ Model \ \|\ $M_X:$ $\alpha=.003$ \ \|\ $M_X:$ $\alpha=.03$ \ \|\ $\alpha_3$:
$\alpha=.003$ \|\ $\alpha_3:$ $\alpha=.03$
 \crthick
Non-Susy $SU(5):$ $N_h=1$
\|\ $M_X>1.1\times 10^{15}GeV$ \| $M_X>3.6\times 10^{15}GeV$ \| $\alpha_3>.140$
\| $\alpha_3>.180$  \nr
Non-Susy $SU(5):$ $N_h=2$
\|\ $M_X>1.1\times 10^{15}GeV$ \| $M_X>3.6\times 10^{15}GeV$ \| $\alpha_3>.153$
\| $\alpha_3>.203$  \nr
Susy $SU(5)$
\|\ $M_X>1.5\times 10^{15}GeV$ \| $M_X>4.7\times 10^{15}GeV$ \| $\alpha_3>.088$
\| $\alpha_3>.100$  \nr
Susy $SU(5)\times U(1)$
\|\ $M_X>1.0\times 10^{15}GeV$ \| $M_X>3.2\times 10^{15}GeV$ \| see graph \|
see graph \endtable}

\hfill}
\smallskip
\parindent=0pt
\centerline{
{\it Table 3} - Limits from dimension six proton decay:
$\tau(p\rightarrow e^+\pi^0)>5.5\times 10^{32}yr$}

\vfill
\break

\vskip 2cm

{\hfill
{\begintable
\ Input Values \ \|\ $\st$ \ \|\ $m_t$ \ \|\ $\ae$ \|\ $m_h$ \|\ $\mu$ \|\ $X$
 \crthick
$X_{max}$
\|\ ~~$0.2337$ \| ~~$92\GeV$ \| ~~$1/128.1$ \|
 $1\TeV$ \| $1\TeV$ \| $195.0$ \nr
central $\st$,$m_t$
\|\ ~~$0.2328$ \| ~~$120\GeV$ \|
 ~~$1/128.1$ \| $1\TeV$ \| $1\TeV$ \| $190.1$ \nr
central $\ae$
\|\ ~~$0.2337$ \| ~~$92\GeV$ \| ~~$1/127.9$ \|
 $1\TeV$ \| $1\TeV$ \| $194.7$ \nr
$X_{central}$
\|\ ~~$0.2328$ \| ~~$120\GeV$ \| ~~$1/127.9$ \|
 $1\TeV$ \| $1\TeV$ \| $189.8$ \nr
$X_{min}$
\|\ ~~$0.2319$ \| ~~$147\GeV$ \| ~~$1/127.7$ \|
 $1\TeV$ \| $1\TeV$ \| $184.6$ \nr
$X_{max}$: $\mu,m_h=500\GeV$
\|\ ~~$0.2337$ \| ~~$92\GeV$ \| ~~$1/128.1$ \|
 $500\GeV$ \| $500\GeV$ \| $192.4$ \endtable}

\hfill}
\smallskip
\parindent=0pt
\centerline{
{\it Table 4} - Sensitivity of X to various inputs:}
\vskip1pt\centerline{$\mu ,m_h = 1$ TeV unless otherwise stated.}

\vfill
\break

\footatend\bigskip\bigskip\bigskip\immediate\closeout\rfile\writestoppt
\baselineskip=14pt\centerline{{\bf References}}\bigskip{\frenchspacing%
\parindent=20pt\escapechar=` \input refs.tmp\vfill\eject}\nonfrenchspacing
\listfigs
\bye